\documentclass[prl,%
twocolumn,%
tightenlines,%
superscriptaddress,showpacs,nofootinbib,amsfonts,amsmath]{revtex4}
\usepackage{bm}

\begin{document}

\title{Gravitational radiation from inspiralling compact binaries\\
completed at the third post-Newtonian order}

\author{Luc Blanchet} \email{blanchet@iap.fr}
\affiliation{${\mathcal{G}}{\mathbb{R}}
\varepsilon{\mathbb{C}}{\mathcal{O}}$, FRE 2435-CNRS, Institut
d'Astrophysique de Paris, 98$^{\text{bis}}$ boulevard Arago, F-75014
Paris, France}

\author{Thibault Damour}
\email{damour@ihes.fr}
\affiliation{Institut des Hautes \'Etudes Scientifiques,
35 route de Chartres, F-91440 Bures-sur-Yvette, France}

\author{Gilles \surname{Esposito-Far\`ese}} \email{gef@iap.fr}
\affiliation{${\mathcal{G}}{\mathbb{R}}
\varepsilon{\mathbb{C}}{\mathcal{O}}$, FRE 2435-CNRS, Institut
d'Astrophysique de Paris, 98$^{\text{bis}}$ boulevard Arago, F-75014
Paris, France}

\author{Bala R. Iyer} \email{bri@rri.res.in} \affiliation{Raman
Research Institute, Bangalore 560 080, India}

\date{June 3, 2004}

\begin{abstract}
The gravitational radiation from point particle binaries is computed
at the third post-Newtonian (3PN) approximation of general
relativity. Three previously introduced ambiguity parameters, coming
from the Hadamard self-field regularization of the 3PN source-type
mass quadrupole moment, are consistently determined by means of
dimensional regularization, and proved to have the values
$\xi=-9871/9240$, $\kappa=0$ and $\zeta=-7/33$. These results complete
the derivation of the general relativistic prediction for compact
binary inspiral up to 3.5PN order, and should be of use for searching
and deciphering the signals in the current network of gravitational
wave detectors.
\end{abstract}

\maketitle

Astrophysical systems known as inspiralling compact binaries (ICBs)
--- two neutron stars or black holes driven into coalescence by
emission of gravitational radiation --- are prominent observable
sources for the gravitational wave observatories LIGO and VIRGO. The
appropriate theoretical description of ICBs is by two structureless
point-particles, characterized solely by their masses $m_1$ or $m_2$
(and possibly their spins), and moving on a quasi-circular orbit.
Strategies to detect and analyze the very weak signals from compact
binary inspiral involve matched filtering of a set of accurate
theoretical template waveforms against the output of the detectors.
Several analyses \cite{3mn,PW95,BCV03} have shown that, in order to
get sufficiently accurate theoretical templates, one must include
high-order post-Newtonian effects, up to the third post-Newtonian
approximation (3PN, or $\sim 1/c^6$, where $c$ denotes the speed of
light), or even better the 3.5PN $\sim 1/c^7$ one. To date, the
templates have been completed through 2.5PN order (for both the phase
\cite{BDIWW95} and amplitude \cite{BIWW96}), and the specific effects
of gravitational wave tails, and tails generated by tails themselves,
have been added up to the 3.5PN order \cite{B98tail}.

Up to now the 3PN-accurate radiation field has only been incompletely
determined. Previous work at 3PN order showed the appearance of
``\textit{ambiguity parameters}'', due to an incompleteness of the
Hadamard regularization (HR) employed for curing the infinite self
field of point particles. By ambiguity parameter we mean an arbitrary
dimensionless coefficient whose value cannot be fixed within HR. In
the binary's 3PN Arnowitt-Deser-Misner Hamiltonian \cite{JaraS98}
there initially appeared two ambiguities, the ``kinetic'' ambiguity
$\omega_k$ and the ``static'' one $\omega_s$, while in the 3PN
equations of motion in harmonic coordinates \cite{BF00} there appeared
a single ambiguity parameter $\lambda$ turning out to be equivalent to
$\omega_s$. The kinetic ambiguity could be resolved by imposing the
global Poincar\'e invariance of the formalism
\cite{BF00,DJSpoinc}. The ADM Hamiltonian and harmonic-coordinates
equations of motion have been shown to yield completely equivalent
results \cite{DJSequiv,ABF01}. More recent work using
\textit{dimensional regularization} (DR) finally determined the static
ambiguity to the value $\omega_s=0$ \cite{DJSdim} or, equivalently,
$\lambda=-1987/3080$ \cite{BDE04}. The same result was also obtained
by means of a surface-integral approach \cite{IFA01}.

We are concerned here with the problem of the binary's 3PN radiation
field (beyond the Newtonian quadrupole formalism), for which
\textit{three} ambiguity parameters, $\xi$, $\kappa$, $\zeta$, have been
shown to appear, coming {}from the HR of the source-type mass quadrupole
moment $I_{ij}$ of point particle binaries at 3PN order \cite{BIJ02}.
The terms corresponding to these ambiguities are given as follows
(see Eq. (10.25) in \cite{BIJ02}),
\begin{eqnarray}
\Delta I_{ij}[\xi,\,\kappa,\,\zeta]
&\equiv&\frac{44}{3}\frac{G_N^2m_1^3}{c^6}\biggl[\left(\xi + \kappa
\frac{m_1+m_2}{m_1}\right) y_1^{\langle i}a_1^{j\rangle}\nonumber\\
&&+\,\zeta
v_1^{\langle i}v_1^{j\rangle}\biggr] + 1\leftrightarrow 2\,.
\label{amb}
\end{eqnarray}
Here, $G_N$ is Newton's constant, the factor $1/c^6$ indicates the 3PN
approximation, and $\mathbf{y}_1$, $\mathbf{v}_1$, $\mathbf{a}_1$
denote the first particle's position, velocity and acceleration. The
symbol $1\leftrightarrow 2$ means the same terms but with all
particles' labels exchanged; the brackets $\langle\rangle$ surrounding
indices refer to the symmetric-trace-free (STF) projection.

In this Letter we present, for the first time, the values of the
parameters $\xi$, $\kappa$ and $\zeta$, and we outline their
derivation using DR (our detailed investigation will be reported in
separate papers \cite{BImult,BDIzeta,BDEIdr}). The main strategy is to
express both the HR and DR results in terms of their ``core'' part,
obtained by applying the so-called ``pure-Hadamard-Schwartz'' (pHS)
regularization. (Following the definition of \cite{BDE04}, the pHS
regularization is a specific, minimal Hadamard-type regularization of
integrals, together with a minimal treatment of ``contact''
ambiguities, and the use of Schwartz distributional derivatives.) The
first step of our calculation \cite{BImult} is to relate the final HR
3PN quadrupole moment, for general orbits, to its pHS part:
\begin{eqnarray}
I_{ij}^{(\mathrm{HR})}[r_1',r_2';\xi,\kappa,\zeta]&=&
I_{ij}^{(\mathrm{pHS})}[r_1',r_2']\nonumber\\
&&+\Delta
I_{ij}[\xi+\hbox{$\frac{1}{22}$},\kappa,\zeta+
\hbox{$\frac{9}{110}$}]\,.
\label{IijH}
\end{eqnarray}
Here, the left-hand side (L.H.S.) denotes the (non-circular
generalization of the) result of \cite{BIJ02}, while the right-hand
side (R.H.S.) contains both the ``core'' pHS quadrupole moment, and the
effect of adding the ambiguities (with some numerical shifts coming
from the difference between the hybrid Hadamard-type regularization
scheme used in \cite{BIJ02} and the pHS one). The pHS part is free of
ambiguity parameters but depends on the specific regularization length
scales $r_1'$ and $r_2'$ introduced in the harmonic-coordinates
equations of motion~\cite{BF00}.

The next step is to derive the multipole moments of an isolated
(slowly-moving) source in $d$ spatial dimensions in order to apply DR
\cite{BDEIdr}. The Einstein field equations in $d+1$ space-time
dimensions are ``relaxed'' by means of the condition of harmonic
coordinates, $\partial_\nu h^{\mu\nu}=0$, where the gravitational field
variable is defined by
$h^{\mu\nu}\equiv\sqrt{g}\,g^{\mu\nu}-\eta^{\mu\nu}$, with
$g^{\mu\nu}$ being the inverse and $-g$ the determinant of the usual
covariant metric, and with $\eta^{\mu\nu}$ the Minkowski metric (in
Minkowskian coordinates). Then,
\begin{equation}\label{EE}
\Box_\eta\,h^{\mu\nu} = \frac{16\pi G}{c^4} \,g \,T^{\mu\nu} +
\Lambda^{\mu\nu}[h] \equiv \frac{16\pi G}{c^4}\,\tau^{\mu\nu}\,,
\end{equation}
where $\Box_\eta$ denotes the flat space-time d'Alembertian operator,
$T^{\mu\nu}$ the matter stress-energy tensor, $\Lambda^{\mu\nu}$
the effective gravitational source term (non-linearly depending on
$h^{\rho\sigma}$ and its space-time derivatives), and
$\tau^{\mu\nu}$ the total stress-energy pseudo tensor of the matter
and gravitational fields. $G$ is related to the usual three-dimensional
  Newton's constant $G_N$ by
$G=G_N\,\ell_0^{d-3}$, where $\ell_0$ denotes an arbitrary length
scale. We have obtained the mass and current multipole moments ($I_L$
and $J_L$) of an arbitrary post-Newtonian source, generalizing the
three-dimensional expressions derived in \cite{B98mult} to any $d$
dimensions. The moments $I_L$ and $J_L$ parametrize the linearized
approximation in the multipolar-post-Minkowskian metric
\textit{exterior} to an isolated source \cite{BD86}. In the case of
the mass-type moments we find
\begin{eqnarray}
I_L(t)&=&\frac{d-1}{2(d-2)}\,\mathrm{FP}_B\int
d^d\mathbf{x}\, \left( \frac{\vert\mathbf{x}\vert}{r_0}\right)^B
\biggl\{\hat{x}_L\,\Sigma_{[\ell]}\nonumber\\
&&-\frac{4(d+2\ell-2)}{c^2(d+\ell-2)(d+2\ell)}
\,\hat x_{aL}\,\Sigma_{[\ell+1]a}^{(1)}\nonumber\\
&&+\frac{2(d+2\ell-2)}{c^4(d+\ell-1)(d+\ell-2)(d+2\ell+2)}
\nonumber\\
&&\times\hat{x}_{abL}
\,\Sigma_{[\ell+2]ab}^{(2)}\biggr\}(\mathbf{x},t)\,,
\label{IL}
\end{eqnarray}
where $L\equiv i_1\cdots i_\ell$ is a multi-index composed of $\ell$
spatial indices ($\ell\geq 2$ is the multipolar order), $\hat{x}_L$ is
the STF part of the product of $\ell$ spatial vectors [\textit{i.e.},
$\hat{x}_L \equiv \mathrm{STF}(x_{i_1}\cdots x_{i_\ell})$], and the
time derivatives are denoted by a superscript $(n)$. The integrand in
(\ref{IL}) is made out of the source densities $\Sigma \,c^2\equiv
2[(d-2)\overline{\tau}^{00}+\overline{\tau}^{ss}]/(d-1)$,
$\Sigma_a\,c\equiv\overline{\tau}^{a0}$ and
$\Sigma_{ab}\equiv\overline{\tau}^{ab}$, built from the formal
\textit{post-Newtonian} expansion, denoted $\overline{\tau}^{\mu\nu}$,
of the pseudo tensor $\tau^{\mu\nu}$. For any of these source
densities the subscript $[\ell]$ denotes the infinite
post-Newtonian-type expansion $\Sigma_{[\ell]}(\mathbf{x},t) \equiv
\sum_{k=0}^{+\infty}\alpha_\ell^k(\vert\mathbf{x}
\vert/c)^{2k}\Sigma^{(2k)}(\mathbf{x},t)$, where the coefficients are
related to the Eulerian $\Gamma$-function by
$\alpha_\ell^k=\Gamma(\frac{d}{2}+\ell)/[2^{2k}k!\,\Gamma
(\frac{d}{2}+\ell+k)]$. The expression (\ref{IL}) involves a
regularization factor $(\vert\mathbf{x}\vert/ r_0)^B$, where
$B\in\mathbb{C}$ and $r_0$ is a separate ``infra-red'' (IR) length
scale, and a particular process of taking the finite part
($\mathrm{FP}_B$), which constitutes the appropriate $d$-dimensional
generalization of the finite part process used in \cite{B98mult} to
treat the IR divergencies of the PN-expanded multipole moments (linked
to the region $\vert\mathrm{x}\vert\rightarrow +\infty$).

Since the source densities $\Sigma$, $\Sigma_a$ and $\Sigma_{ab}$
depend on the post-Newtonian expansion of the metric,
$\overline{h}^{\mu\nu}$, they are obviously to be iterated in a
post-Newtonian way in order to obtain a useful result. At the 3PN
order it is convenient to parametrize the moments by means of the
explicit retarded potentials $V$, $V_a$, $\hat{W}_{ab}$, $\hat{R}_a$
and $\hat{X}$, introduced when $d=3$ in \cite{BF00} and generalized to
$d$ dimensions in \cite{BDE04}. Starting from the matter source
densities $\sigma c^2\equiv 2[(d-2)T^{00}+T^{ss}]/(d-1)$, $\sigma_a
c\equiv T^{a0}$ and $\sigma_{ab}\equiv T^{ab}$, we first define the
``linear'' potentials $V=\Box^{-1}_R[-4\pi G \sigma]$ and
$V_a=\Box^{-1}_R[-4\pi G \sigma_a]$, where $\Box^{-1}_R$ is the usual
flat space-time retarded operator. The linear potentials are then used
to construct higher ``non-linear'' potentials, such as
$\hat{W}_{ab}=\Box^{-1}_R[-4\pi G
(\sigma_{ab}-\frac{1}{d-2}\,\delta_{ab}\,
\sigma_{ss})-\frac{d-1}{2(d-2)}\,\partial_a V\partial_b V]$. The
retardations in the latter potentials are systematically expanded to
the required PN order. At Newtonian order the expression
(\ref{IL}) reduces to the standard result $I_L=\int
d^d\mathbf{x}\,\rho\,\hat{x}_L+\mathcal{O}(c^{-2})$ with
$\rho=T^{00}/c^2$.

We have used Eq. (\ref{IL}) to compute the \textit{difference} between
the DR result and the pHS one \cite{BDEIdr}. As in the work on
equations of motion \cite{DJSdim,BDE04}, we find that the ambiguities
arise solely from the terms in the integration regions near the
particles ($r_1=\vert\mathbf{x}-\mathbf{y}_1\vert \to 0$ or
$r_2=\vert\mathbf{x}-\mathbf{y}_2\vert \to 0$) that give rise to poles
$\propto 1/\varepsilon$ (where $\varepsilon\equiv d-3$), corresponding
in 3 dimensions to logarithmic ultraviolet (UV) divergencies. We have
verified (thanks to the appropriate definition of the finite part
process $\mathrm{FP}_B$) that the IR region
($\vert\mathrm{x}\vert\rightarrow +\infty$) did not contribute to the
difference DR $-$ pHS. The ``compact-support'' terms in the integrand
of (\ref{IL}), \textit{i.e.}, the terms proportional to the matter
source densities $\sigma$, $\sigma_a$ and $\sigma_{ab}$, were also
found not to contribute to the difference (thanks to the definition of
``contact'' terms in pHS \cite{BDE04}). We are therefore left with
evaluating the difference linked with the computation of the
\textit{non-compact} terms in the \textit{local} expansion of the
integrand in (\ref{IL}) near the singularities (\textit{i.e.},
$r_1\rightarrow 0$ and $r_2\rightarrow 0$) that produce poles in $d$
dimensions. Let us denote by $F^{(d)}$ the non-compact part of the
integrand of the $d$-dimensional multipole moment (\ref{IL})
(including the appropriate multipolar factors such as $\hat{x}_L$),
that is to say we write the non-compact part of $I_L$ as the integral
$\int d^d\mathbf{x}\,F^{(d)}$, extended, say, over two small domains
$0< r_1< \mathcal{R}_1$, $0< r_2< \mathcal{R}_2$. [At this stage we
can set $B=0$ in $F^{(d)}$, and remove the $\mathrm{FP}_B$
prescription.] We write the expansion of $F^{(d)}$ when
$r_1\rightarrow 0$, up to any order $N$, in the form (\textit{cf.}
\cite{BDE04})
\begin{equation}\label{Fd}
F^{(d)}(\mathbf{x})=\sum_{p,q}r_1^{p+q\varepsilon}{}_1f_{p,q}^{
(\varepsilon)}(\mathbf{n}_1)+o(r_1^N)\,,
\end{equation}
where $p,q$ are relative integers, and
$\mathbf{n}_1\equiv (\mathbf{x}-\mathbf{y}_1)/r_1$.

In practice the ${}_1f_{p,q}^{ (\varepsilon)}$'s are computed by
specializing the general expressions of the non-linear retarded
potentials $V$, $V_a$, $\hat{W}_{ab}$, $\cdots$ (valid for general
extended sources) to the point particles case in $d$ dimensions. The
matter source density reads $\sigma =
\tilde{\mu}_1(t)\,\delta^{(d)}(\mathbf{x}-\mathbf{y}_1)
+1\leftrightarrow 2$, where $\delta^{(d)}$ is Dirac's delta-function
in $d$ dimensions, and $\tilde{\mu}_1(t)$ denotes a certain function
of time, coming from the standard prescription for point particles in
general relativity, and which is computed in an iterative
post-Newtonian way. The function $\tilde{\mu}_1$ depends on the
potentials $V$, $V_a$, $\cdots$, evaluated at the location of the
singular points following the rules of DR, \textit{i.e.}, by invoking
analytic continuation in $d\in\mathbb{C}$. We PN expand the
(time-symmetric) propagators. For instance, at the 1PN order, we use
$\Box^{-1}=\Delta^{-1}+\frac{1}{c^2}\Delta^{-2}\partial_t^2
+\mathcal{O}(c^{-4})$, which yields the solution
$V=G\,\tilde{\mu}_1\,\tilde{k}\,r_1^{2-d}+\partial_t^2\bigl[
G\,\tilde{\mu}_1\,\tilde{k}\,r_1^{4-d}/2c^2(4-d)\bigr]
+1\leftrightarrow 2+\mathcal{O}(c^{-4})$, where we denote
$\tilde{k}\equiv\Gamma (\frac{d-2}{2})/\pi^{\frac{d-2}{2}}$.
Proceeding further, we insert the previous solution for $V$ into the
quadratic part of the source term for $\hat{W}_{ab}$ (whose structure
is $\sim\partial V\partial V$), expand it when $r_1\rightarrow 0$, and
then integrate \textit{term by term} in order to find the local
expansion (when $r_1\rightarrow 0$) of the corresponding solution,
using the integration formula $\Delta^{-1}
\bigl[\hat{n}_1^L\,r_1^\alpha\bigr] =
\hat{n}_1^L\,r_1^{\alpha+2}/\bigl[(\alpha-\ell+2)(\alpha+\ell+d)
\bigr]$. We generate by this method a \textit{particular} solution of
the Poisson-like equation we want to solve, and we added the
supplementary homogeneous solution defined in Ref.~\cite{BDE04}, whose
programs we have re-used for the present work.

The difference $\mathcal{D}I$ between the DR evaluation of the
($d$-dimensional) local integral $\int d^d\mathbf{x}
\,F^{(d)}(\mathbf{x})$, and its corresponding, three-dimensional pHS
evaluation, \textit{i.e.}, the ``partie finie'' $\mathrm{Pf}_{s_1,s_2}
\int d^3\mathbf{x} \,F^{(d=3)}(\mathbf{x})$, is expressible in terms of
the expansion coefficients of (\ref{Fd}) as (see also \cite{DJSdim})
\begin{eqnarray}
\mathcal{D}I[s_1,s_2;\varepsilon,\ell_0]&=&
\frac{\Omega_{2+\varepsilon}}{\varepsilon} \sum_q
\left[\frac{1}{q+1}+\varepsilon \ln s_1\right]\nonumber\\
&&\times\langle\,{}_1f_{-3,q}^{
(\varepsilon)}(\mathbf{n}_1)\,\rangle+1\leftrightarrow 2 +
\mathcal{O}(\varepsilon)\,,\hspace{5mm}
\label{diff}
\end{eqnarray}
where the brackets denote the angular average over the unit sphere in
$3+\varepsilon$ dimensions (with total volume
$\Omega_{2+\varepsilon}$) centered on the singularity
$\mathbf{y}_1$. The L.H.S. depends both on the regularization length
scales $s_1, s_2$ of the Hadamard partie finie, and on the DR
regularization characteristics, $\varepsilon=d-3$ and $\ell_0$.

With this definition, the dimensional regularization of the 3PN
quadrupole moment (indices $L\equiv ij$) is obtained as the sum of its
``pure-Hadamard-Schwartz'' part, and of a ``difference'' computed
according to Eq.~(\ref{diff}):
$I_{ij}^{(\mathrm{DR})}[\varepsilon,\ell_0] =
I_{ij}^{(\mathrm{pHS})}[s_1,s_2]+\mathcal{D}I_{ij}[
s_1,s_2;\varepsilon,\ell_0].$ At this stage a check of the result is
that the HR scales $s_1$ and $s_2$ cancel out between the two terms in
the R.H.S., so that $I_{ij}^{(\mathrm{DR})}$ depends only on
$\varepsilon$ and $\ell_0$ [its dependence on $\varepsilon$ is of the
form of a simple pole followed by a finite part,
$a_{-1}\,\varepsilon^{-1}+a_0+\mathcal{O}(\varepsilon)$]. Because of
this independence from $s_1$, $s_2$, we can express our result in
terms of the constants $r'_1$ and $r'_2$ used as fiducial scales in the
final result of~\cite{BIJ02,BImult}, see Eq. (\ref{IijH}). We can
therefore write the DR result as
\begin{equation}\label{Iijdr'}
I_{ij}^{(\mathrm{DR})}[\varepsilon,\ell_0] =
I_{ij}^{(\mathrm{pHS})}[r'_1,r'_2]+\mathcal{D}I_{ij}[
r'_1,r'_2;\varepsilon,\ell_0]\,.
\end{equation}

Let us now impose the \textit{physical equivalence} between the DR
result (\ref{Iijdr'}) and the corresponding HR result (\ref{IijH}). In
doing this identification, we must remember that the ``bare particle
positions'', $\mathbf{y}_1^\mathrm{bare}$ and
$\mathbf{y}_2^\mathrm{bare}$, entering the DR result differ from their
Hadamard counterparts (used in \cite{BF00,BIJ02,BImult}) by some
\textit{shifts}, which were uniquely determined in \cite{BDE04}, and
denoted there
$\bm{\xi}_1(r_1';\varepsilon,\ell_0)$ and
$\bm{\xi}_2(r_2';\varepsilon,\ell_0)$ (see Eqs. (1.13) and
(6.41)--(6.43) in \cite{BDE04}). In the present work, we will denote
them by $\bm{\eta}_1$ and $\bm{\eta}_2$ in order to avoid any
confusion with the ambiguity parameter $\xi$. In other words, we
impose the equivalence
\begin{equation}\label{shift}
I_{ij}^{(\mathrm{HR})}[r_1',r_2';\xi,\kappa,\zeta] =
\lim_{\varepsilon\rightarrow 0}\left[
I_{ij}^{(\mathrm{DR})}[\varepsilon,\ell_0]+\delta_{\bm{\eta}(r_1',r_2';
\varepsilon,\ell_0)}I_{ij}\right]\,,
\end{equation}
in which
$\delta_{\bm{\eta}}I_{ij}=2\,m_1\,y_1^{<i}\eta_1^{j>}+1\leftrightarrow
2$ denotes the total change induced on the quadrupole moment by the
latter shifts. We find that the poles $\sim 1/\varepsilon$ separately
present in the two terms in the brackets of (\ref{shift}) cancel, so
that the physical (``dressed'') DR quadrupole moment is \textit{
finite} and given by the limit shown in (\ref{shift}).

Finally, by inserting the expressions of the DR and HR results given
respectively by (\ref{Iijdr'}) and (\ref{IijH}) into
Eq. (\ref{shift}), and by removing the pHS part which is common to
both results, we obtain a relation for the ambiguity part $\Delta
I_{ij}$ of the quadrupole moment in (\ref{IijH}) in terms of known
quantities,
\begin{eqnarray}
\Delta
I_{ij}[\xi+\hbox{$\frac{1}{22}$},\kappa,\zeta+\hbox{$\frac{9}{110}$}]
&=&\lim_{\varepsilon\rightarrow
0}\bigl[\mathcal{D}I_{ij}[r_1',r_2';\varepsilon,\ell_0]\nonumber\\
&&+\,\delta_{\bm{\eta}(r_1',r_2'; \varepsilon,\ell_0)}I_{ij}\bigr]\,.
\label{equamb}
\end{eqnarray}
The apparent dependence of the R.H.S. on $r_1'$, $r_2'$, $\varepsilon$
and $\ell_0$, is checked to cancel out. Equation (\ref{equamb}) then
gives \textit{three} equations for the three unknowns
$\xi,\kappa,\zeta$, thereby yielding the central result of this work:
\begin{equation}\label{res}
\xi=-\frac{9871}{9240}\,,~~\kappa=0\,,~~\zeta=-\frac{7}{33}\,,
\end{equation}
which finally provides an unambiguous determination of the 3PN
radiation field by DR.

We have been able to perform several checks of our calculation. First
of all, we have also obtained $\zeta$ by considering the limiting
physical situation where the mass of one of the particles is exactly
zero (say $m_2=0$), and the other particle moves with uniform
velocity. Computing the quadrupole moment of a \textit{boosted}
Schwarzschild black hole at 3PN order and comparing the result with
$I_{ij}^{(\mathrm{HR})}$ in the limit $m_2=0$ we recover exactly the
same value for $\zeta$ \cite{BDIzeta}. This agreement is a direct
verification of the global Poincar\'e invariance of the wave
generation formalism, and a check that DR automatically preserves this
invariance (as it did in the context of the equations of motion
\cite{DJSdim,BDE04}). Secondly, the fact that the value of $\kappa$ is
zero has been checked by showing that there are no corresponding
dangerously divergent ``diagrams'' (in the sense of \cite{Dgef96}).
Finally, we have computed the DR of the mass \textit{dipole} moment,
$I_i^{(\mathrm{DR})}$, following the same method and using the same
computer programs, and found that it differs by exactly the same
shifting of the world-lines, $\bm{\eta}_1(r_1';\varepsilon,\ell_0)$ and
$\bm{\eta}_2(r_2';\varepsilon,\ell_0)$, as in Eq. (\ref{shift}),
\textit{i.e.}, $\delta_{\bm{\eta}}I_i=m_1\,\eta_1^i+1\leftrightarrow
2$, from the 3PN center-of-mass position associated with the equations
of motion in harmonic coordinates as given by Eq. (4.5) of
\cite{ABF01}.

The determination of the values (\ref{res}) completes the problem of the
general relativistic prediction for the templates of ICBs up to 3PN
order (and actually up to 3.5PN as the corresponding ``tail terms''
have already been determined \cite{B98tail}). The relevant combination
of the parameters (\ref{res}) entering the 3PN energy flux in the case
of circular orbits, namely $\theta$ \cite{BIJ02}, is now fixed to
\begin{equation}
\theta\equiv\xi+2\kappa+\zeta=-\frac{11831}{9240}\,.
\label{theta}\end{equation}
Numerically, $\theta\simeq -1.28041$. The orbital phase of compact
binaries, in the adiabatic inspiral regime (\textit{i.e.}, evolving by
radiation reaction), involves at 3PN order a linear combination of
$\theta$ and of the equation-of-motion related parameter $\lambda$
\cite{BFIJ02}, which is determined as
\begin{equation}
\hat{\theta}\equiv \theta-\frac{7}{3}\lambda=\frac{1039}{4620}\,.
\label{thetahat}\end{equation}
The fact that the numerical value of this parameter is quite small,
$\hat{\theta}\simeq 0.22489$, indicates, following
measurement-accuracy analyses \cite{BCV03}, that the 3PN (or better
3.5PN) order should provide an excellent approximation for both the
on-line search and the subsequent off-line analysis of gravitational
wave signals from ICBs in the LIGO and VIRGO detectors.

\acknowledgments
L.B. and B.R.I. thank IFCPAR for its support.


\begin{thebibliography}{39}
\expandafter\ifx\csname natexlab\endcsname\relax\def\natexlab#1{#1}\fi
\expandafter\ifx\csname bibnamefont\endcsname\relax
\def\bibnamefont#1{#1}\fi \expandafter\ifx\csname
bibfnamefont\endcsname\relax \def\bibfnamefont#1{#1}\fi
\expandafter\ifx\csname citenamefont\endcsname\relax
\def\citenamefont#1{#1}\fi \expandafter\ifx\csname url\endcsname\relax
\def\url#1{\texttt{#1}}\fi \expandafter\ifx\csname
urlprefix\endcsname\relax\def\urlprefix{URL }\fi
\providecommand{\bibinfo}[2]{#2
\providecommand{\eprint}[2][]{\url{#2}}}

\bibitem[{\citenamefont{Cutler
et~al.}(1993{\natexlab{a}})\citenamefont{Cutler, Apostolatos,
Bildsten, Finn, Flanagan, Kennefick, Markovic, Ori, Poisson, Sussman
et~al.}}]{3mn}
\bibinfo{author}{\bibfnamefont{C.}~\bibnamefont{Cutler}},
\bibnamefont{et~al.}, \bibinfo{journal}{Phys. Rev. Lett.}
\textbf{\bibinfo{volume}{70}}, \bibinfo{pages}{2984}
(\bibinfo{year}{1993}{\natexlab{a}});
\bibinfo{author}{\bibfnamefont{C.}~\bibnamefont{Cutler}}
\bibnamefont{and}
\bibinfo{author}{\bibfnamefont{E.E.}~\bibnamefont{Flanagan}},
\bibinfo{journal}{Phys. Rev. D} \textbf{\bibinfo{volume}{49}},
\bibinfo{pages}{2658} (\bibinfo{year}{1994});
\bibinfo{author}{\bibfnamefont{H.}~\bibnamefont{Tagoshi}}
\bibnamefont{and}
\bibinfo{author}{\bibfnamefont{T.}~\bibnamefont{Nakamura}},
\bibinfo{journal}{Phys. Rev. D} \textbf{\bibinfo{volume}{49}},
\bibinfo{pages}{4016} (\bibinfo{year}{1994});
\bibinfo{author}{\bibfnamefont{E.}~\bibnamefont{Poisson}},
\bibinfo{journal}{Phys. Rev. D} \textbf{\bibinfo{volume}{52}},
\bibinfo{pages}{5719} (\bibinfo{year}{1995}), \bibinfo{note}{erratum
Phys. Rev. D \textbf{55}, 7980 (1997)}.

\bibitem[{\citenamefont{Poisson and Will}(1995)}]{PW95}
\bibinfo{author}{\bibfnamefont{E.}~\bibnamefont{Poisson}}
\bibnamefont{and}
\bibinfo{author}{\bibfnamefont{C.M.}~\bibnamefont{Will}},
\bibinfo{journal}{Phys. Rev. D} \textbf{\bibinfo{volume}{52}},
\bibinfo{pages}{848} (\bibinfo{year}{1995});
\bibinfo{author}{\bibfnamefont{A.}~\bibnamefont{Krol\`ak}},
\bibinfo{author}{\bibfnamefont{K.D.}~\bibnamefont{Kokkotas}}
\bibnamefont{and}
\bibinfo{author}{\bibfnamefont{G.}~\bibnamefont{Sch\"afer}},
\bibinfo{journal}{Phys. Rev. D} \textbf{\bibinfo{volume}{52}},
\bibinfo{pages}{2089} (\bibinfo{year}{1995});
\bibinfo{author}{\bibfnamefont{T.}~\bibnamefont{Damour}},
\bibinfo{author}{\bibfnamefont{B.R.}~\bibnamefont{Iyer}}
\bibnamefont{and}
\bibinfo{author}{\bibfnamefont{B.S.}~\bibnamefont{Sathyaprakash}},
\bibinfo{journal}{Phys. Rev. D} \textbf{\bibinfo{volume}{57}},
\bibinfo{pages}{885} (\bibinfo{year}{1998}); \textit{ibid.}
\textbf{\bibinfo{volume}{62}}, \bibinfo{pages}{084036}
(\bibinfo{year}{2000}).

\bibitem[{\citenamefont{Buonanno
et~al.}(2003{\natexlab{a}})\citenamefont{Buonanno, Chen, and
Vallisneri}}]{BCV03}
\bibinfo{author}{\bibfnamefont{A.}~\bibnamefont{Buonanno}},
\bibinfo{author}{\bibfnamefont{Y.}~\bibnamefont{Chen}}
\bibnamefont{and}
\bibinfo{author}{\bibfnamefont{M.}~\bibnamefont{Vallisneri}},
\bibinfo{journal}{Phys. Rev. D} \textbf{\bibinfo{volume}{67}},
\bibinfo{pages}{024016} (\bibinfo{year}{2003}{\natexlab{a}});
\bibinfo{journal}{Phys. Rev. D} \textbf{\bibinfo{volume}{67}},
\bibinfo{pages}{104025} (\bibinfo{year}{2003}{\natexlab{b}});
\bibinfo{author}{\bibfnamefont{T.}~\bibnamefont{Damour}},
\bibinfo{author}{\bibfnamefont{B.R.}~\bibnamefont{Iyer}},
\bibinfo{author}{\bibfnamefont{P.}~\bibnamefont{Jaranowski}}
\bibnamefont{and}
\bibinfo{author}{\bibfnamefont{B.S.}~\bibnamefont{Sathyaprakash}},
\bibinfo{journal}{Phys. Rev. D}
\textbf{\bibinfo{volume}{67}}, \bibinfo{pages}{064028}
(\bibinfo{year}{2003}).

\bibitem[{\citenamefont{Blanchet
et~al.}(1995{\natexlab{a}})\citenamefont{Blanchet, Damour, Iyer, Will,
and Wiseman}}]{BDIWW95}
\bibinfo{author}{\bibfnamefont{L.}~\bibnamefont{Blanchet}},
\bibinfo{author}{\bibfnamefont{T.}~\bibnamefont{Damour}},
\bibinfo{author}{\bibfnamefont{B.R.}~\bibnamefont{Iyer}},
\bibinfo{author}{\bibfnamefont{C.M.}~\bibnamefont{Will}}
\bibnamefont{and}
\bibinfo{author}{\bibfnamefont{A.G.}~\bibnamefont{Wiseman}},
\bibinfo{journal}{Phys. Rev. Lett.}
\textbf{\bibinfo{volume}{74}}, \bibinfo{pages}{3515}
(\bibinfo{year}{1995}{\natexlab{a}});
\bibinfo{author}{\bibfnamefont{L.}~\bibnamefont{Blanchet}},
\bibinfo{author}{\bibfnamefont{T.}~\bibnamefont{Damour}}
\bibnamefont{and}
\bibinfo{author}{\bibfnamefont{B.R.}~\bibnamefont{Iyer}},
\bibinfo{journal}{Phys. Rev. D}
\textbf{\bibinfo{volume}{51}}, \bibinfo{pages}{5360}
(\bibinfo{year}{1995}{\natexlab{b}});
\bibinfo{author}{\bibfnamefont{C.M.}~\bibnamefont{Will}}
\bibnamefont{and}
\bibinfo{author}{\bibfnamefont{A.G.}~\bibnamefont{Wiseman}},
\bibinfo{journal}{Phys. Rev. D} \textbf{\bibinfo{volume}{54}},
\bibinfo{pages}{4813} (\bibinfo{year}{1996});
\bibinfo{author}{\bibfnamefont{L.}~\bibnamefont{Blanchet}},
\bibinfo{journal}{Phys. Rev. D} \textbf{\bibinfo{volume}{54}},
\bibinfo{pages}{1417} (\bibinfo{year}{1996}).

\bibitem[{\citenamefont{Blanchet
et~al.}(1996{\natexlab{a}})\citenamefont{Blanchet, Iyer, Will, and
Wiseman}}]{BIWW96}
\bibinfo{author}{\bibfnamefont{L.}~\bibnamefont{Blanchet}},
\bibinfo{author}{\bibfnamefont{B.R.}~\bibnamefont{Iyer}},
\bibinfo{author}{\bibfnamefont{C.M.}~\bibnamefont{Will}}
\bibnamefont{and}
\bibinfo{author}{\bibfnamefont{A.G.}~\bibnamefont{Wiseman}},
\bibinfo{journal}{Class. Quant. Grav.}
\textbf{\bibinfo{volume}{13}}, \bibinfo{pages}{575}
(\bibinfo{year}{1996});
\bibinfo{author}{\bibfnamefont{K.G.}~\bibnamefont{Arun}},
\bibinfo{author}{\bibfnamefont{L.}~\bibnamefont{Blanchet}},
\bibinfo{author}{\bibfnamefont{B.R.}~\bibnamefont{Iyer}}
\bibnamefont{and}
\bibinfo{author}{\bibfnamefont{M.S.}~\bibnamefont{Qusailah}}
(\bibinfo{year}{2004}),
\bibinfo{journal}{Class. Quant. Grav. submitted},
\eprint{gr-qc/0404085}.

\bibitem[{\citenamefont{Blanchet}(1998{\natexlab{a}})}]{B98tail}
\bibinfo{author}{\bibfnamefont{L.}~\bibnamefont{Blanchet}},
\bibinfo{journal}{Class. Quant. Grav.} \textbf{\bibinfo{volume}{15}},
\bibinfo{pages}{113} (\bibinfo{year}{1998}{\natexlab{a}}).

\bibitem[{\citenamefont{Jaranowski and Sch\"afer}(1998)}]{JaraS98}
\bibinfo{author}{\bibfnamefont{P.}~\bibnamefont{Jaranowski}}
\bibnamefont{and}
\bibinfo{author}{\bibfnamefont{G.}~\bibnamefont{Sch\"afer}},
\bibinfo{journal}{Phys. Rev. D} \textbf{\bibinfo{volume}{57}},
\bibinfo{pages}{7274} (\bibinfo{year}{1998});
\bibinfo{journal}{Phys. Rev. D} \textbf{\bibinfo{volume}{60}},
\bibinfo{pages}{124003} (\bibinfo{year}{1999}).

\bibitem[{\citenamefont{Blanchet and Faye}(2000)}]{BF00}
\bibinfo{author}{\bibfnamefont{L.}~\bibnamefont{Blanchet}}
\bibnamefont{and}
\bibinfo{author}{\bibfnamefont{G.}~\bibnamefont{Faye}},
\bibinfo{journal}{Phys. Lett. A} \textbf{\bibinfo{volume}{271}},
\bibinfo{pages}{58} (\bibinfo{year}{2000});
\bibinfo{journal}{Phys. Rev. D} \textbf{\bibinfo{volume}{63}},
\bibinfo{pages}{062005} (\bibinfo{year}{2001}).

\bibitem[{\citenamefont{Damour et~al.}(2000)\citenamefont{Damour,
Jaranowski, and Sch\"afer}}]{DJSpoinc}
\bibinfo{author}{\bibfnamefont{T.}~\bibnamefont{Damour}},
\bibinfo{author}{\bibfnamefont{P.}~\bibnamefont{Jaranowski}}
\bibnamefont{and}
\bibinfo{author}{\bibfnamefont{G.}~\bibnamefont{Sch\"afer}},
\bibinfo{journal}{Phys. Rev. D} \textbf{\bibinfo{volume}{62}},
\bibinfo{pages}{021501R} (\bibinfo{year}{2000}),
\bibinfo{note}{erratum Phys. Rev. D \textbf{63}, 029903, (2001)}.

\bibitem[{\citenamefont{Damour
et~al.}(2001{\natexlab{a}})\citenamefont{Damour, Jaranowski, and
Sch\"afer}}]{DJSequiv}
\bibinfo{author}{\bibfnamefont{T.}~\bibnamefont{Damour}},
\bibinfo{author}{\bibfnamefont{P.}~\bibnamefont{Jaranowski}}
\bibnamefont{and}
\bibinfo{author}{\bibfnamefont{G.}~\bibnamefont{Sch\"afer}},
\bibinfo{journal}{Phys. Rev. D} \textbf{\bibinfo{volume}{63}},
\bibinfo{pages}{044021} (\bibinfo{year}{2001}{\natexlab{a}}).

\bibitem[{\citenamefont{de~Andrade
et~al.}(2001)\citenamefont{de~Andrade, Blanchet, and Faye}}]{ABF01}
\bibinfo{author}{\bibfnamefont{V.}~\bibnamefont{de~Andrade}},
\bibinfo{author}{\bibfnamefont{L.}~\bibnamefont{Blanchet}}
\bibnamefont{and}
\bibinfo{author}{\bibfnamefont{G.}~\bibnamefont{Faye}},
\bibinfo{journal}{Class. Quantum Grav.}
\textbf{\bibinfo{volume}{18}}, \bibinfo{pages}{753}
(\bibinfo{year}{2001}).

\bibitem[{\citenamefont{Damour
et~al.}(2001{\natexlab{b}})\citenamefont{Damour, Jaranowski, and
Sch\"afer}}]{DJSdim}
\bibinfo{author}{\bibfnamefont{T.}~\bibnamefont{Damour}},
\bibinfo{author}{\bibfnamefont{P.}~\bibnamefont{Jaranowski}}
\bibnamefont{and}
\bibinfo{author}{\bibfnamefont{G.}~\bibnamefont{Sch\"afer}},
\bibinfo{journal}{Phys. Lett. B} \textbf{\bibinfo{volume}{513}},
\bibinfo{pages}{147} (\bibinfo{year}{2001}{\natexlab{b}}).

\bibitem[{\citenamefont{Blanchet et~al.}(2004)\citenamefont{Blanchet,
Damour, and Esposito-Far{\`e}se}}]{BDE04}
\bibinfo{author}{\bibfnamefont{L.}~\bibnamefont{Blanchet}},
\bibinfo{author}{\bibfnamefont{T.}~\bibnamefont{Damour}}
\bibnamefont{and}
\bibinfo{author}{\bibfnamefont{G.}~\bibnamefont{Esposito-Far{\`e}se}}
(\bibinfo{year}{2004}), \bibinfo{note}{Phys. Rev. D in press},
\eprint{gr-qc/0311052}.

\bibitem[{\citenamefont{Itoh et~al.}(2001)\citenamefont{Itoh,
Futamase, and Asada}}]{IFA01}
\bibinfo{author}{\bibfnamefont{Y.}~\bibnamefont{Itoh}},
\bibinfo{author}{\bibfnamefont{T.}~\bibnamefont{Futamase}}
\bibnamefont{and}
\bibinfo{author}{\bibfnamefont{H.}~\bibnamefont{Asada}},
\bibinfo{journal}{Phys. Rev. D} \textbf{\bibinfo{volume}{63}},
\bibinfo{pages}{064038} (\bibinfo{year}{2001});
\bibinfo{author}{\bibfnamefont{Y.}~\bibnamefont{Itoh}}
\bibnamefont{and}
\bibinfo{author}{\bibfnamefont{T.}~\bibnamefont{Futamase}},
\bibinfo{journal}{Phys. Rev. D} \textbf{\bibinfo{volume}{68}},
\bibinfo{pages}{121501} (\bibinfo{year}{2003});
\bibinfo{author}{\bibfnamefont{Y.}~\bibnamefont{Itoh}},
\bibinfo{journal}{Phys. Rev. D} \textbf{\bibinfo{volume}{69}},
\bibinfo{pages}{064018} (\bibinfo{year}{2004}).

\bibitem[{\citenamefont{Blanchet
et~al.}(2002{\natexlab{a}})\citenamefont{Blanchet, Iyer, and
Joguet}}]{BIJ02}
\bibinfo{author}{\bibfnamefont{L.}~\bibnamefont{Blanchet}},
\bibinfo{author}{\bibfnamefont{B.R.}~\bibnamefont{Iyer}}
\bibnamefont{and}
\bibinfo{author}{\bibfnamefont{B.}~\bibnamefont{Joguet}},
\bibinfo{journal}{Phys. Rev. D} \textbf{\bibinfo{volume}{65}},
\bibinfo{pages}{064005} (\bibinfo{year}{2002}{\natexlab{a}}).

\bibitem{BImult}L.~Blanchet and B.R.~Iyer, \textit{Hadamard
regularization of the 3PN wave generation of two point masses}, in
preparation.

\bibitem{BDIzeta}L.~Blanchet, T.~Damour and B.R.~Iyer,
\textit{Multipole moments of a boosted Schwarzschild black hole}, in
preparation.

\bibitem{BDEIdr}L.~Blanchet, T.~Damour, G.~Esposito-Far\`ese, and
B.R.~Iyer, \textit{Dimensional regularization of the 3PN wave
generation of two point masses}, in preparation.

\bibitem[{\citenamefont{Blanchet}(1998{\natexlab{b}})}]{B98mult}
\bibinfo{author}{\bibfnamefont{L.}~\bibnamefont{Blanchet}},
\bibinfo{journal}{Class. Quant. Grav.} \textbf{\bibinfo{volume}{15}},
\bibinfo{pages}{1971} (\bibinfo{year}{1998}{\natexlab{b}});
\bibinfo{journal}{Phys. Rev. D} \textbf{\bibinfo{volume}{51}},
\bibinfo{pages}{2559} (\bibinfo{year}{1995}{\natexlab{b}}).

\bibitem[{\citenamefont{Blanchet and Damour}(1986)}]{BD86}
\bibinfo{author}{\bibfnamefont{L.}~\bibnamefont{Blanchet}}
\bibnamefont{and}
\bibinfo{author}{\bibfnamefont{T.}~\bibnamefont{Damour}},
\bibinfo{journal}{Phil. Trans. R. Soc. London A}
\textbf{\bibinfo{volume}{409}}, \bibinfo{pages}{383}
(\bibinfo{year}{1986}).

\bibitem[{\citenamefont{Damour and
Esposito-Far{\`e}se}(1996)}]{Dgef96}
\bibinfo{author}{\bibfnamefont{T.}~\bibnamefont{Damour}}
\bibnamefont{and}
\bibinfo{author}{\bibfnamefont{G.}~\bibnamefont{Esposito-Far{\`e}se}},
\bibinfo{journal}{Phys. Rev.} \textbf{\bibinfo{volume}{D53}},
\bibinfo{pages}{5541} (\bibinfo{year}{1996}).

\bibitem[{\citenamefont{Blanchet
et~al.}(2002{\natexlab{b}})\citenamefont{Blanchet, Faye, Iyer, and
Joguet}}]{BFIJ02}
\bibinfo{author}{\bibfnamefont{L.}~\bibnamefont{Blanchet}},
\bibinfo{author}{\bibfnamefont{G.}~\bibnamefont{Faye}},
\bibinfo{author}{\bibfnamefont{B.R.}~\bibnamefont{Iyer}}
\bibnamefont{and}
\bibinfo{author}{\bibfnamefont{B.}~\bibnamefont{Joguet}},
\bibinfo{journal}{Phys. Rev.} \textbf{\bibinfo{volume}{D65}},
\bibinfo{pages}{061501} (\bibinfo{year}{2002}{\natexlab{b}}).

\end{thebibliography}

\end{document}